\documentstyle[11pt,newpasp,twoside]{article}
\index{globular clusters}
\index{millisecond pulsars}
\index{low mass x-ray binaries}
\index{magnetic fields}
\index{accretion}
\index{polar cap heating}

\input{psfig.sty}

\def\about{$\sim$}

\def\erg/cm2sec{erg~cm$^{-2}$~s$^{-1}$}  
\def\ergcm2{erg~cm$^{-2}$}

\def\X{$\times$~}

\def\Lx{L$_x$~}

\def\pc3{pc$^{-3}$~}

\def\cm-3{cm{$^{-3}$}~} 
\def\km/s{km~s$^{-1}$~}

\def\Edot{{$\dot{E}$}~}

\def\X{$\times$}
\def\sdage{$P/2\dot P$}

\def\vdisp{$\sigma_{v0}$}

\def\apj{ApJ}
\def\aa{A\&A}

\newcommand{\lsim }{{\lower0.8ex\hbox{$\buildrel <\over\sim$}}}
\newcommand{\gsim }{{\lower0.8ex\hbox{$\buildrel >\over\sim$}}}

\newcommand{\Msun}{\ifmmode {M_{\odot}}\else${M_{\odot}}$\fi~}
\newcommand{\Rsun}{\ifmmode {R_{\odot}}\else${R_{\odot}}$\fi~}
\newcommand{\Lsun}{\ifmmode {L_{\odot}}\else${L_{\odot}}$\fi~}
\newcommand{\mv}{\ifmmode {m_{V}}\else${m_{V}}$\fi}
\newcommand{\Mv}{\ifmmode {M_{V}}\else${M_{V}}$\fi}
\newcommand{\lopt}{\ifmmode L_{opt} \else $~L_{opt}$\fi}
\newcommand{\loglopt}{\ifmmode{\rm log}~L_{opt} \else log$~L_{opt}$\fi}
\newcommand{\lx}{\ifmmode L_x \else $~L_x$\fi}
\newcommand{\loglx}{\ifmmode{\rm log}~L_x \else log$~L_x$\fi}
\newcommand{\cmsq}{\ifmmode{\rm ~cm^{-2}} \else cm$^{-2}$\fi}
\newcommand{\nh}{\ifmmode{\rm N_{H}} \else N$_{H}$\fi}
\newcommand{\fcgs}{\ifmmode {\rm erg~cm}^{-2}~{\rm s}^{-1}\else
erg~cm$^{-2}$~s$^{-1}$\fi} 
\newcommand{\lcgs}{\ifmmode erg~~s^{-1}\else erg~s$^{-1}$\fi}

\reversemarginpar

\begin{document}
\title{{\it Chandra} on Millisecond Pulsars in Globular Clusters}
 \author{J.E. Grindlay, C.O. Heinke and P.D. Edmonds}
\affil{ Harvard-Smithsonian Center for Astrophysics, 60 Garden
  Street, Cambridge, MA~02138, USA}
\author{F. Camilo}
\affil{Columbia Astrophysics Laboratory, Columbia University,
  550 West 120th Street, New York, NY~10027, USA}

\begin{abstract}
We summarize the x-ray properties of the complete samples of 
millisecond pulsars (MSPs) detected in our {\it Chandra} observations 
of the globular clusters 47~Tuc and NGC~6397. The 47~Tuc
MSPs are predominantly soft sources suggestive of thermal emission from
the neutron star polar cap and have x-ray luminosities in a
surprisingly narrow range (\Lx \about1--4 \X 10$^{30}$ \lcgs).  
The single MSP in NGC~6397 is both hard and apparently extended, probably 
due to shocked hot gas evaporating from its main sequence companion.
In contrast to MSPs in the field and the cluster M28, which 
show correlation between x-ray 
luminosity and spindown luminosity \Lx $\propto$ \Edot$^{\beta}$ with 
$\beta$ \about1 - 1.4, the 47~Tuc (and NGC~6397) sample 
display a relatively tight correlation with $\beta = 0.5\pm0.15$.
The correlations of \Lx vs. \sdage~ and light cylinder 
magnetic field values are also different. 
It is possible the magnetic field configuration has 
been altered (by episodic accretion) for old MSPs in dense cluster cores. 

\end{abstract}

\section{Introduction}
X-ray studies of millisecond pulsars (MSPs) can constrain 
fundamental properties of their emission regions and, when 
combined with radio timing studies, their underlying neutron 
stars (NSs). In globular clusters 
both MSPs and low mass x-ray binaries (LMXBs), 
their likely progenitors,  are significantly enhanced (per unit mass) 
over their values in the galactic disk by stellar and binary 
interactions. The dense cluster (core) environment needed for 
their excess formation may also alter their evolution. Thus 
cluster vs. field MSPs, as studied in x-rays and radio, can 
constrain intrinsic vs. extrinsic (evolutionary) properties of 
these oldest NS systems.

We have conducted a deep {\it Chandra} survey for MSPs as well as  
quiescent LMXBs and cataclysmic variables (CVs)  
in the globular clusters 47~Tuc (Grindlay et al. 2001a; GHE01a) 
and NGC~6397 (Grindlay et al. 2001b; GHE01b). The full details 
of the MSP survey are given in Grindlay et al. (2001c; GCH01). 
Here we present the highlights of this study, focusing on just 
the x-ray properties of the 16 MSPs with radio timing 
positions in 47~Tuc (Freire et al. 2001a, Freire 2001) and the one 
in NGC~6397 (D'Amico et al. 2001; DPM) as well as their comparison 
with the field MSP population 
(cf. Becker \& Trumper 1997, 1999; BT97, BT99). 
We defer to the full paper the discussion 
of the  total MSP populations and 
spatial distributions, which probe cluster dynamics.

\section{X-ray Colors and Emission Models}
The 47~Tuc MSPs were found initially (GHE01a) to be soft sources. 
In GCH01 we give the detected
counts in 3 bands: softcts (0.2--1\,keV), mediumcts (1--2\,keV) and
hardcts (2--8\,keV) for each of the 14 resolved MSPs, with counts for
47~Tuc-G and -I (unresolved) estimated. 
From these bands, we form the hardness ratios
HR1 = mediumcts/softcts and HR2 = hardcts/mediumcts and plot the MSPs,
with counting statistics errors, in the color-color diagram shown in
Figure~1 (left).  The MSP colors are clustered in a relatively
narrow range of HR1 and HR2 with 47~Tuc-J  clearly harder, as was 
evident in the Xcolor distributions in GHE01a.

Using the PIMMS tool, we construct values of 
HR1 and HR2 for 3 simple models:  thermal
bremsstrahlung (TB), blackbody (BB) and power law (PL), with index
values (kT or photon index) given in the caption of
Figure~1 (left). The observed range of HR1-HR2
is roughly consistent with TB spectra with kT \about 1\,keV, BB spectra
with kT \about 0.2--0.3\,keV (except for 47~Tuc-J)
or PL spectra with photon index \about3. The weighted mean colors for
all but 47~Tuc-J are consistent with a BB spectrum with kT \about0.22\,keV, 
giving  x-ray luminosities 
\Lx(0.5-2.5keV) \about1-4 \X 10$^{30}$ \lcgs 
and thus mean bolometric L$_{x-bol}$ = 2.6 \X 10$^{30}$ \lcgs. 

The x-ray colors rule out TB models 
(surrounding column densities inconsistent 
with the MSP dispersion measures; DM) and PL fits (spectral indices 
implausible). Simple BB fits for  L$_{x-bol}$ give emission radii 
of only \about0.1km whereas H (or He)-atmosphere models 
(Rajagopal \& Romani 1996) typically give 
temperatures reduced (from BB) by a factor 
of \about2  and thus radii increased to \about0.4km. 
Either case  suggests soft x-ray emission from a region 
smaller than the entire polar cap, as predicted in recent models of 
Harding \& Muslimov (2001) for polar cap heating.  Although the 
3.2s temporal resolution of {\it Chandra}-ACIS prevents a pulsation analysis, 
the small thermal emission area suggests the emission would be
pulsed, with a sinusoidal pulse shape appropriate to the fractional
visibility of the isotropically radiating thermal polar cap. In
contrast, the narrower pulse duty cycles of \about10\% for some field
MSPs (and one in the globular cluster M28; BT99) 
are probably due to non-thermal beamed emission.

\begin{figure}[t]
\hspace*{-0.3in}
\centerline{\psfig{file=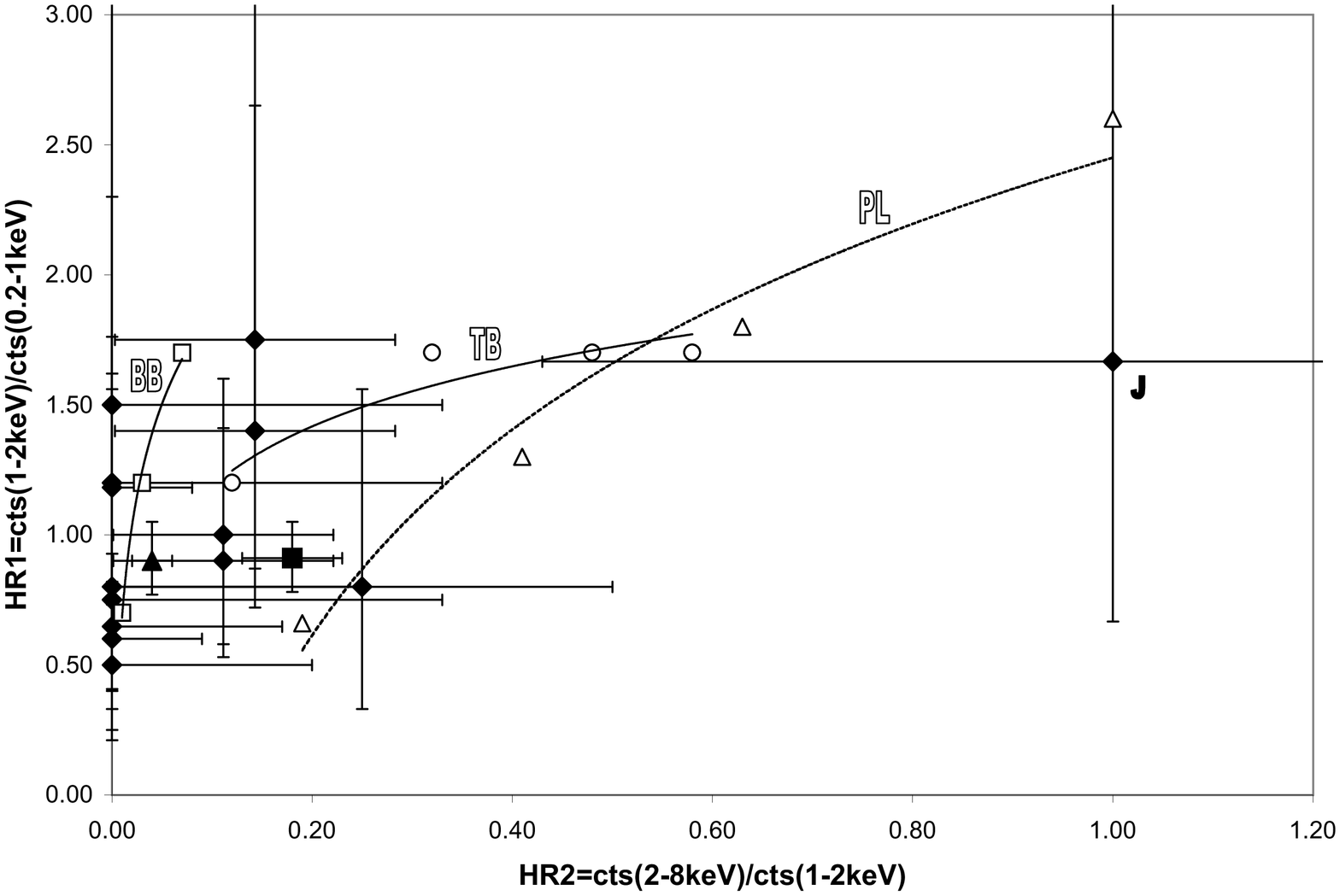,width=2.3in,height=2.5in,angle=-90.}\hspace*{0.3in}\psfig{file=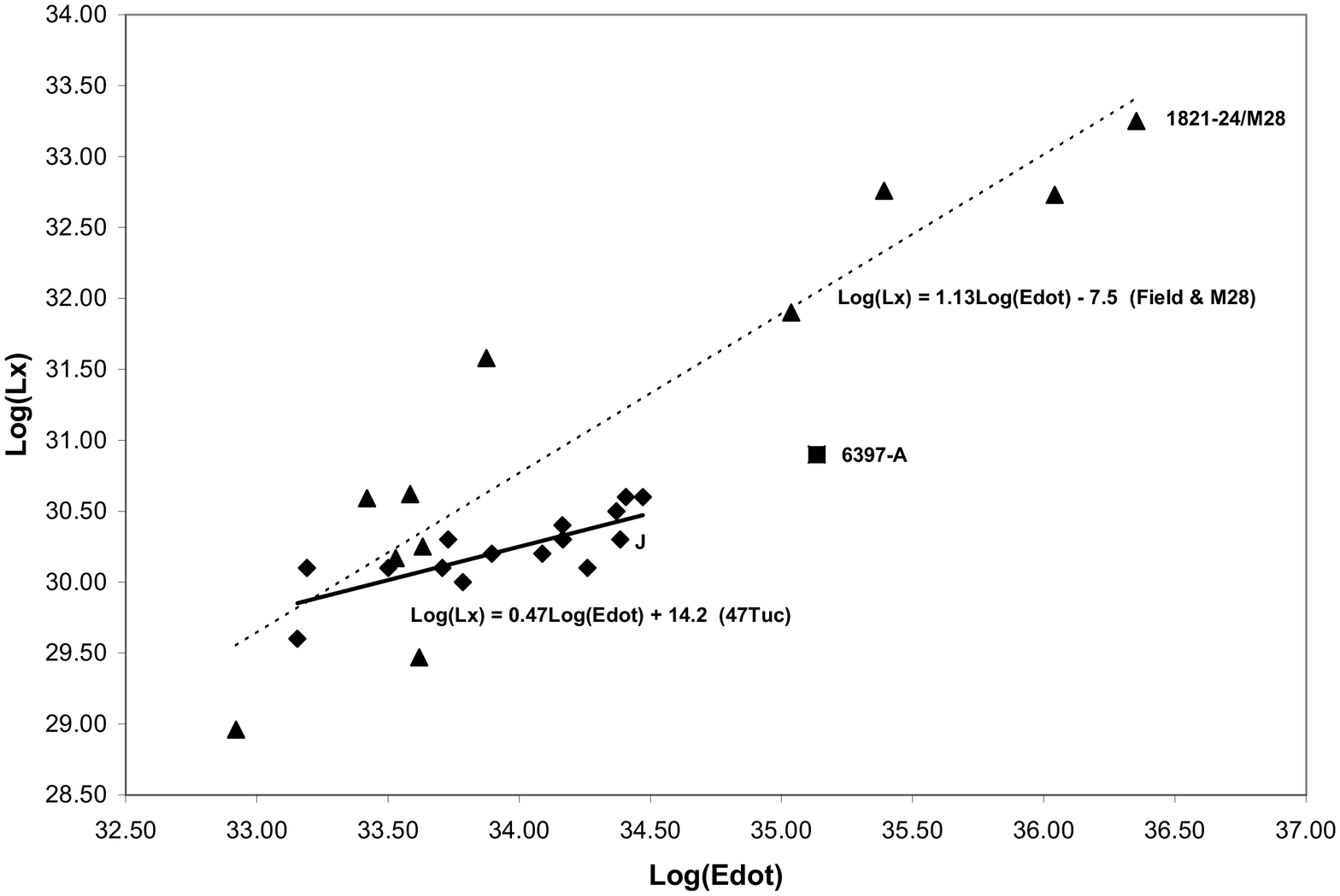,width=2.3in,height=2.5in,angle=-90.}}\\
\vspace{-0.2in}
\caption{{\it Left:} X-ray color-color diagram for MSPs in 47~Tuc
(solid diamonds) vs. tracks for models with parameters changing from
lower left to upper right: blackbody (BB; open squares), with kT = 0.2,
0.25, 0.3\,keV; power law (PL; open triangles), with photon index
$\alpha$ = 3, 2, 1.5, 1; and thermal bremsstrahlung (TB; open circles),
with kT = 1, 2, 3, 6\,keV.  Weighted mean colors for the MSPs in 47~Tuc
with and without 47~Tuc-J shown as the solid
square and triangle, respectively.
%
{\it Right:} \Lx vs. \Edot for MSPs in 47~Tuc
(diamonds) and
field (triangles).  MSP in NGC~6397 (box) 
is labeled as is J in 47~Tuc.} 
\end{figure}

\section{ X-ray vs. Pulsar Spin Properties}
A key question for this rich {\it Chandra} 
dataset is the correlation of x-ray luminosity and pulsar spindown
luminosity $\dot E$, which is found for field MSPs (with much more
uncertain distances) to scale as \Lx(0.1-2.4keV) \about 10$^{-3}$ 
\Edot (BT97) and with a possibly steeper logarithmic slope (1.4) for 
\Lx in the 2-10keV band (Possenti et al. 2001; PCC). We derive 
instrinsic period derivatives, $\dot P_i$,  
corrected for the cluster acceleration by estimating the 
3D positions of each MSP in the cluster from the observed DM value 
and the observed hot gas and thus electron density 
in the cluster (Freire et al. 2001b) 
and then subtracting the cluster acceleration using 
a King model with cluster parameters derived by 
Meylan \& Mayor (1986). Using a standard NS moment of inertia
$I=10^{45}$\,g\,cm$^2$, we then derive \Edot  = $4\pi^2 I \dot P_i/P^3$
for each MSP and plot them 
vs.  \Lx(0.5-2.5keV) in Figure~1 (right). Uncertainties in the  
\Edot values are typically 0.2--0.5 in the log 
but are not shown for clarity; uncertainties in log(L$_x$) 
are typically 0.2, and extrapolating to the {\it ROSAT} band, 0.1-2.4keV, 
would increase log(L$_x$) only by 0.1. For comparison with 
47~Tuc, we plot the MSP in NGC~6397 (GHE01b), for which 
the \Edot uncertainty is small, and updated values (cf. GCH01) 
for the 10 field MSPs previously detected in x-rays as well as 
in the globular cluster M28.

Whereas the MSPs in the field and M28 show (Figure 1, right) a correlation 
log\Lx(0.1-2.4keV) = (1.13$\pm0.15$)log\Edot - 7.5$\pm5$, the MSPs in
47~Tuc appear to have a weaker dependence:  log\Lx(0.5-2.5keV) =
($0.47\pm0.10$)log\Edot + $14.2\pm3.4$ for the nominal cluster model
with central velocity dispersion 
\vdisp = 11.6\,km\,s$^{-1}$, where the errors ($\pm$1$\sigma$) in 
both correlations are due to just the scatter in the points. 
Allowing for uncertainties in the 
cluster model and distance gives slope $0.48\pm0.21$ and 
intercept $13.8\pm7.5$. Including the errors for the \Edot
values estimated for the 47~Tuc MSPs, but with the
approximation that unequal errors (on $\dot E$) are simply averaged
(which biases the slope to steeper values, since the unequal errors are
much larger for smaller values of $\dot E$), increases the 
logarithmic slope to
$\beta = 0.62\pm0.29$ and offset to $9.0\pm10.8$. The best (median) 
estimate for the 47~Tuc MSPs alone is thus $\beta$ \about$0.55\pm0.2$.  
Apart from the uncertain detection of 47~Tuc-C (GCH01), 
the MSPs in 47~Tuc have \Lx(0.5--2.5\,keV) values within 
a factor of \about4 despite a range of
\about25 in $\dot E$. Figure 1 (right) shows that 6397-A, the MSP in NGC~6397, 
is consistent with the \Lx - \Edot correlation shown by the 
47~Tuc MSPs. Including 6397-A in the 
fit yields $\beta = 0.50\pm0.08$ (scatter only) or 0.55$\pm0.15$ (with
averaged $\pm1\sigma$ errors for the 47~Tuc \Edot values).

The smaller $\beta$ values for 47~Tuc and NGC~6397 may be due to
the different formation histories and possibly different physical
parameters of their MSPs vs. the field objects. In Figure~2 (left) we plot
\Lx vs. spindown ages, $P/2\dot P_i$, for MSPs in the field (and M28) vs. 
the 47~Tuc and NGC~6397 sample. Error 
bars on the age parameter are not shown, for clarity, but are typically 
$\sim ^{+0.3}_{-0.1}$ in the log and primarily due to the 47~Tuc 
acceleration model. Despite the uncertainties, the correlation
is striking: the field MSPs show a declining \Lx with ``age,'' and the
47~Tuc MSPs appear to fall on this trend but 
with a flatter \Lx vs. age slope.  However, spindown ages 
correspond only approximately to actual ages, and while these estimates are
consistent with the formation of most 47~Tuc MSPs early in the cluster
history, this would not by itself provide an explanation for the
different \Lx vs. \Edot  correlations found for MSPs 
in 47~Tuc and NGC~6397 vs. the field. 

\begin{figure}[t]
\hspace*{-0.3in}
\centerline{\psfig{file=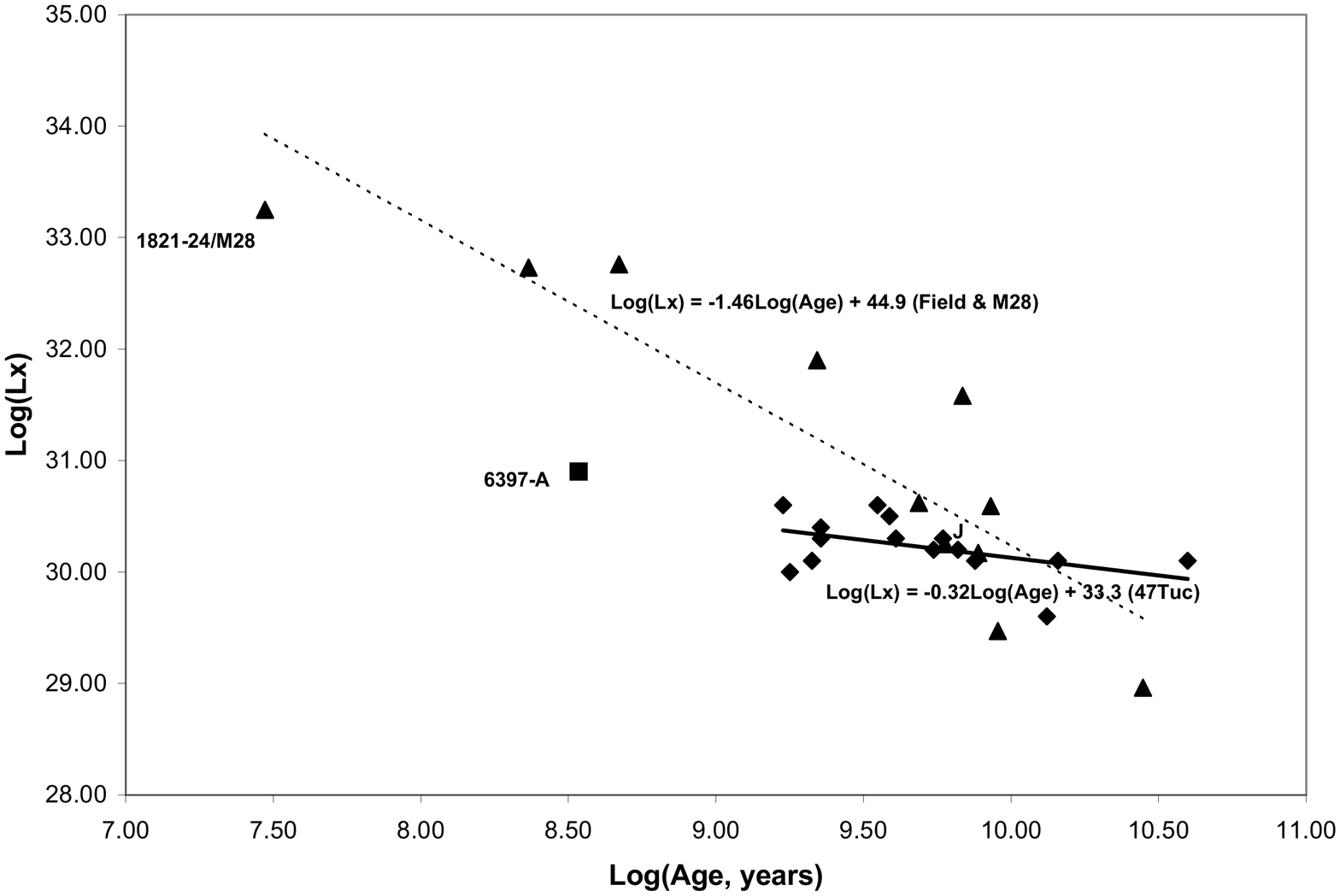,width=2.3in,height=2.5in,angle=-90.}\hspace*{0.3in}\psfig{file=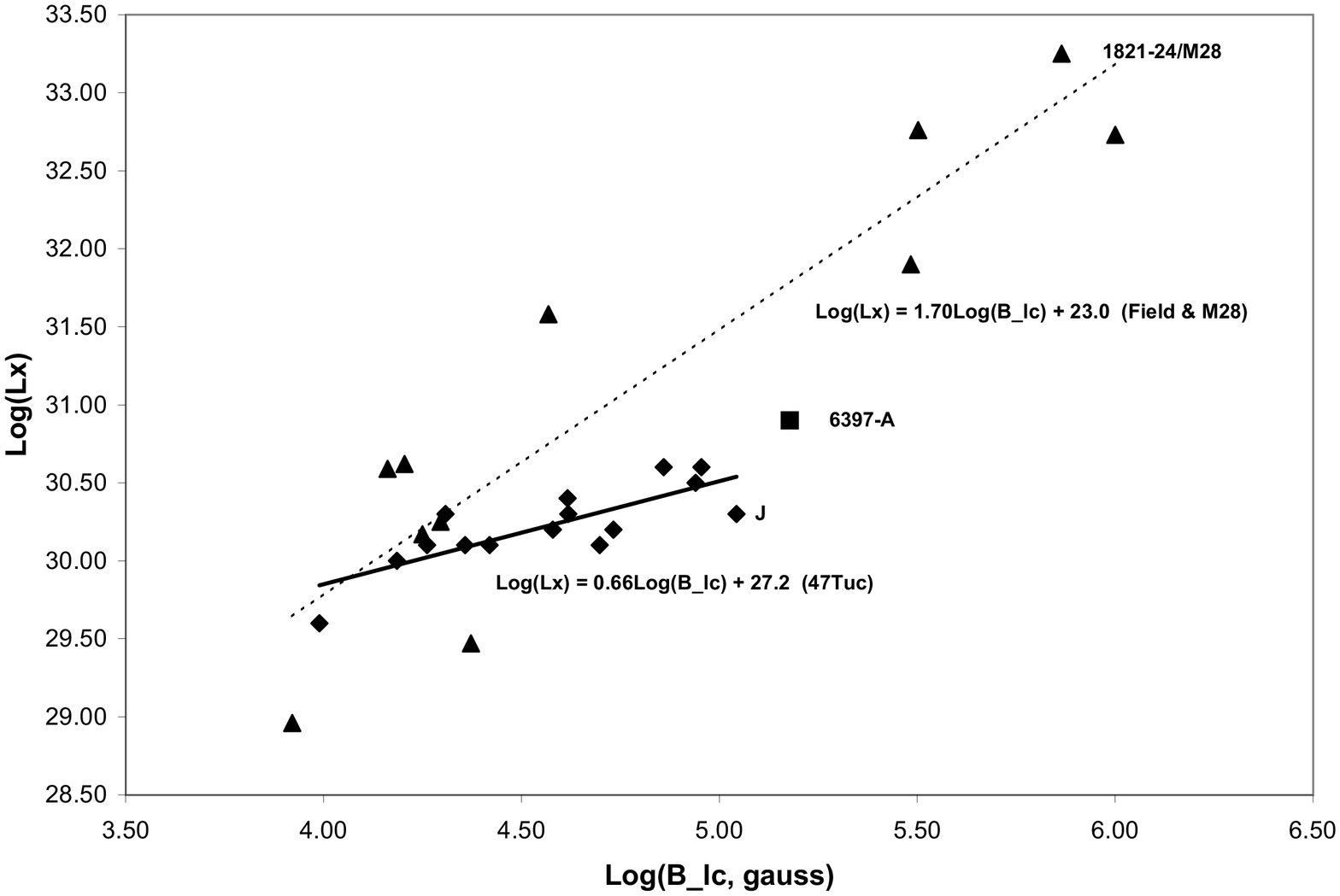,width=2.3in,height=2.5in,angle=-90.}}
\vspace{-0.2in}
\caption{{\it Left:} \Lx vs. characteristic age
($P/2\dot P_i$) for MSPs in 47~Tuc (diamonds) and field (triangles).
MSP in NGC~6397 (box) is labeled as is J in 47~Tuc. 
%
{\it Right:} \Lx vs. $B_{\rm lc}$, 
inferred magnetic
field strength at the light cylinder ($r_{\rm lc}=c P/2\pi$) for MSPs
in 47~Tuc (diamonds) and field (triangles).  MSP in NGC~6397 (box) is 
labeled as is J in 47~Tuc.  Values for MSPs in field and M28 
from data given in GCH01, with $B_{\rm lc}$ derived assuming standard
$1/r^3$ dependence of dipole field (see text).}
\end{figure}

Another possible physical difference between the 47~Tuc and field MSPs 
might be magnetic field strength at the light cylinder (at which the 
corotation speed equals $c$), since this likely affects the relative 
importance of non-thermal (magnetospheric) emission. 
For an assumed dipole field, this is given by 
$B_{\rm lc} = 9.35 \times 10^5 \dot P_{i}^{1/2} P_{\rm msec}^{-5/2}$ 
G, with $\dot P_{i}$ in units of 10$^{-20}$.
In Figure~2 (right) we plot \Lx vs. $B_{\rm lc}$. 
Again, considered as a homogeneous group, the MSPs in the field (and M28) 
lie on a steeper slope than the 47~Tuc MSPs. In support of the hypothesis 
that $B_{\rm lc}$, rather than the field 
$B_{\rm surf} =  3.2 \times 10^{19}$ ($P\dot{P}$)$^{1/2}$\,G 
at the NS surface, is more closely 
related to \Lx is the fact that the correlation of \Lx with $B_{\rm surf}$ 
is less defined and with larger scatter, and with logarithmic slopes 
differing even more: 0.05$\pm0.27$ for the 47~Tuc MSPs 
vs. 2.80$\pm0.99$ for the field MSPs. We note that 3 out of the 4 
MSPs with $B_{\rm lc} \ga 10^{5.5}$\,G display x-ray emission that is seemingly
magnetospheric, with the nature of emission from the 4th (the eclipsing
MSP B1957+20) indeterminate (BT99; Takahashi et al.~2001).  Conversely,
field pulsars with $B_{\rm lc} < 10^{5}$\,G have x-ray emission that is
typically either thermal or of indeterminate character (BT99).
Considering the small numbers of such pulsars studied, and that most of
them have highly uncertain distances, it
seems possible that field pulsars with $B_{\rm lc} \la 10^{5}$\,G may
show an L$_x$--$B_{\rm lc}$ trend that is fairly flat and roughly
consistent with the better determined relation for the 47~Tuc MSPs. 
However for this interpretation to hold, a few field MSPs with 
relatively well determined distances (e.g. 0437$-$4715 and 1744$-$1134; cf. 
GCH01) must be accounted for and the even larger 
deviations of 0751+1807 and 1024$-$0719 from the 47~Tuc correlation line 
would require factor \gsim3 adjustments in these MSP distances. 

\section{Conclusions}
The MSPs in 47~Tuc are primarily very soft x-ray sources,
consistent with thermal emission from the pulsar polar caps. 
The MSPs in both 47~Tuc and NGC~6397 
seem to have a less efficient conversion of
rotational spindown energy ($\dot E$) into soft x-rays (L$_x$) than
most field MSPs, even those with correspondingly low values 
of their magnetic field at the light cylinder. 
For \Lx $\propto$ \Edot$^{\beta}$, the 47~Tuc-NGC~6397 
samples are fit by $\beta
= 0.5\pm0.15$ whereas the updated (GCH01) 
field (and M28) sample are consistent with the
value $\beta \sim 1 - 1.4$ found previously (BT97; PCC and
references therein).

The soft colors for the 47~Tuc MSPs and the extended emission from
6397-A (GCH01) indicate a lack of beamed magnetospheric emission,  
which may suggest  
they have different  $B_{\rm lc}$ values or configurations. 
We speculate their B fields may have a  multipole field geometry 
(and thus lower dipole component) and be less efficient particle 
accelerators. Unlike MSPs in
the field, those in dense cluster cores have a possibility of being
driven back into contact and an accretion phase, as a re-cycled LMXB
(from a MSP)!~ Renewed accretion (and 6397-A is presently close to
filling its Roche lobe; cf.\ DPM) would likely continue the B-field
burial process thought to be responsible (e.g.\ Romani 1990) for field
decay from the $\ga10^{11}$\,G fields at NS birth to the $\la10^9$\,G
values typical of MSPs.  The 47~Tuc MSPs have spent their  $\tau
\ga 1$\,Gyr lifetime in a dense ($n \sim 10^5\,{\rm pc}^{-3}$) cluster
core, where they undergo scattering interactions with both single stars 
and other binaries.  Such scattering causes the binding energy $x$ of a
hard binary to secularly increase. 
Thus, while angular momentum transfer to the secondary acts to detach
the binary, scattering events tend to drive it back toward contact.

A complication in this picture is that since MSPs are extremely hard
binaries, the
secular increase in $x$ is largely the result of infrequent strong
scattering events, which are likely to eject the 
binary from the cluster core or even the cluster.  
Nevertheless, it appears plausible that a typical old ($\tau
\ga 1$\,Gyr) MSP in a \emph{dense} cluster core might undergo one or
more MSP-LMXB transformation cycles. Thus a few percent of the MSPs at any
one time may be in (or near) this recurrent LMXB phase. This might
explain the puzzling luminous qLMXBs X5 and X7 (GHE01a and Heinke et
al.~2002) in 47~Tuc:  they might be recently ($\sim10^{5-6}$\,yr)
detached LMXBs, in which their underlying MSP nature is hidden by their
evaporating (from MSP-driven winds?) fossil disks, thus explaining their 
lack of any detectable low-level accretion signatures. 

The younger system in NGC~6397 has the advantages of having been
recently scattered out of a still higher density (\about10$^6$
pc$^{-3}$) core collapsed cluster core  and possibly having exchanged
its companion (Ferraro et al. 2001, GHE01b). 
Either or both would likely have restored an accretion
phase. Thus 6397-A need not be just ``born,'' as suggested by 
Ferraro et al. (2001); it
may instead have just been reborn.  In contrast, the MSP in M28 is both
\about10\X\ younger, single and in a lower density core and so is unlikely to
have gone through a renewed accretion phase. 

\acknowledgments
We thank H. Cohn and P. Lugger for discussions of MSP recycling, 
described in GCH01. 
This work was supported in part by NASA
grants GO0-1098A and HST-AR-09199.01-A (JG) and NAG5-9095 (FC).




\end{document}